\documentclass[10pt]{article} 
\usepackage[accepted]{tmlr}

\usepackage{amsmath,amsfonts,bm}

\def\eqref#1{equation~\ref{#1}}

\def\1{\bm{1}}

\DeclareMathAlphabet{\mathsfit}{\encodingdefault}{\sfdefault}{m}{sl}
\SetMathAlphabet{\mathsfit}{bold}{\encodingdefault}{\sfdefault}{bx}{n}

\usepackage{hyperref}
\usepackage{url}
\usepackage{microtype}
\usepackage{enumitem}
\usepackage{algorithm} 
\usepackage[noend]{algpseudocode} 
\usepackage{graphicx}
\usepackage{tabularx}
\usepackage{longtable}
\usepackage{wrapfig}
\usepackage{svg}
\usepackage{subfigure}
\usepackage{booktabs} 

\title{ProtiGeno: a prokaryotic short gene finder using protein language models}

\author{\name Tony Tu \email ttu32@gatech.edu \\
      \addr School of Electrical and Computer Engineering \\
      Georgia Institute of Technology
      \AND
      \name Gautham Krishna \email gramalaxmi3@gatech.edu  \\
      \addr School of Biological Sciences \\
      Georgia Institute of Technology
      \AND
      \name Amirali Aghazadeh \email amiralia@gatech.edu\\
      \addr School of Electrical and Computer Engineering  \\
      Georgia Institute of Technology}

\def\month{07}  
\def\year{2023} 

\begin{document}

\maketitle

\begin{abstract}
Prokaryotic gene prediction plays an important role in understanding the biology of organisms and their function with applications in medicine and biotechnology. Although the current gene finders are highly sensitive in finding long genes, their sensitivity decreases noticeably in finding shorter genes ($<$180 nts). The culprit is insufficient annotated gene data to identify distinguishing features in short open reading frames (ORFs). We develop a deep learning-based method called ProtiGeno, specifically targeting short prokaryotic genes using a protein language model trained on millions of evolved proteins. In systematic large-scale experiments on 4,288 prokaryotic genomes, we demonstrate that ProtiGeno predicts short coding and noncoding genes with higher accuracy and recall than the current state-of-the-art gene finders. We discuss the predictive features of ProtiGeno and possible limitations by visualizing the three-dimensional structure of the predicted short genes. Data, codes, and models are available at  \href{https://github.com/tonytu16/protigeno}{https://github.com/tonytu16/protigeno}. 
\end{abstract}

\section{Introduction}
\label{submission}

\begin{figure*}[htb]
    \centering
    \includegraphics[width=1.0\textwidth]{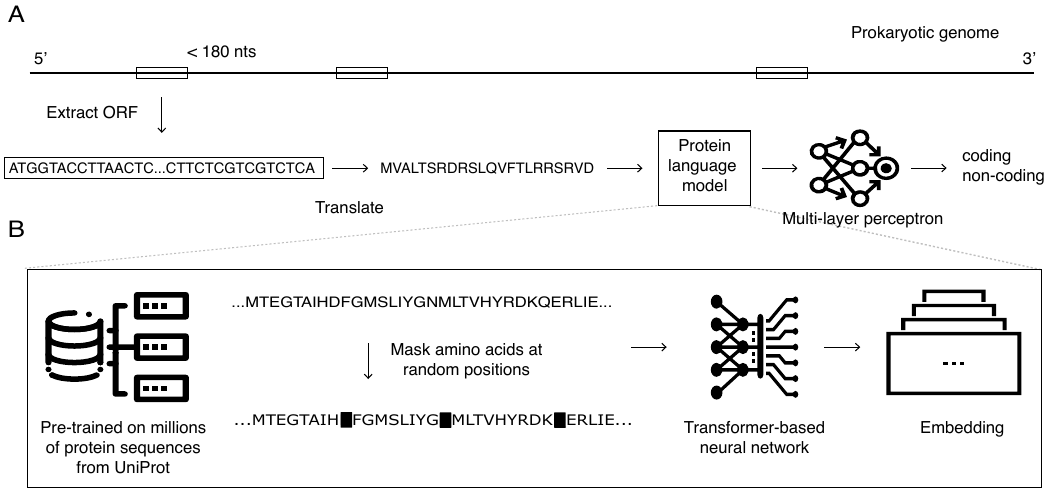}
    \caption{A) Schematic of our ProtiGeno method. ProtiGeno extracts ORFs from a prokaryotic genome and translates them to amino acid sequences. It then finds a 1280-dimensional embedding vector using a protein language model. ProtiGeno passes the embedding vector through a multi-layer perceptron and obtains a binary output indicating whether the input sequence is protein coding or noncoding. B) Overview of protein language model. We use a protein language model called ESM-1b from Meta AI~\cite{rives2021biological}. ESM-1b is trained on the UR50/S dataset ~\cite{suzek2015uniref} from the UniProt database using a variant of the masked language modeling technique on protein sequences. The training process involves randomly masking a certain percentage of amino acids for each input sequence and asking the model to predict the missing amino acids. This enables the model to learn representations that capture meaningful biological properties of the amino acid sequence, such as their secondary and tertiary structures. 
}
    \label{fig:fig1}
\end{figure*}

Advances in low-cost and high-throughput sequencing technologies have led to a deluge of genomic data from prokaryotic organisms. Automatic gene annotation tools are crucial for describing the biological functions of genes and ultimately understanding these organisms. Most of the current prokaryotic gene finders belong to one of the families of Glimmer~\cite{delcher1999improved}, GeneMarkS~\cite{besemer2001genemarks}, and Prodigal~\cite{hyatt2010prodigal}, all of which are based on Markov models ~\cite{rabiner1986introduction} equipped with various biologically-inspired heuristics (e.g., GC content and the Shine–Dalgarno sequence). While these methods are highly sensitive for finding long protein-coding genes, their sensitivity (recall rate) drops drastically for finding short coding genes ($<$180 nts) (see Figure~\ref{fig:gms_recall_vs_length}). One reason behind this performance drop is the lack of sufficiently rich genomic features that could be gleaned from the Markov models~\cite{zhang2017gene}. Departing from the conventional Markov-based methods, \emph{ab inito} machine learning methods have been recently developed for gene finding. Balrog~\cite{sommer2021balrog}, for example, trains a temporal convolutional deep neural network on a diverse collection of high-throughput prokaryotic organisms. These supervised models capture a richer class of genomic signatures for gene prediction than Markov models. They also alleviate the need for retraining for each target genome, allowing for a universal gene prediction model. However, similar to Markov-based models, these machine learning models are primarily developed for annotating long genes. As a result, their sensitivity for short genes remains low due to a lack of annotated short gene sequences in the RefSeq database ~\cite{nielsen2005large, skovgaard2001total}. This lack of information from short genes renders the machine learning methods ill-equipped to identify genes belonging to many underrepresented species, resulting in missing gene types~\cite{warren2010missing}. It stands to reason that we need to borrow statistical power from independent, unsupervised biological data resources to compensate for the scarcity of short genes and complex distinguishing features within them to develop highly sensitive models for gene annotation.

Parallel to the recent advances in gene prediction, Transformer-based deep learning models ~\cite{rao2020transformer} trained on the amino acid sequences of the proteins evolved throughout millions of years of evolution and stored in protein databases such as UniProt ~\cite{uniprot2015uniprot} have revolutionized protein structure prediction~\cite{jumper2021highly,sapoval2022current}. In particular, masked protein language models have demonstrated success in fast and accurate protein structure prediction without explicitly training to learn the rules of protein folding~\cite{lin2022language}. This power of protein language models in predicting the secondary and tertiary structure and biological functions of many proteins motivates the main intellectual motivation behind our approach to gene prediction. Herein, we hypothesize that it is feasible to build a sensitive and universal short gene finder by training a classifier on the output embedding vector of a protein language model once input with the translated sequence of a short open reading frame. Having this goal in mind, we developed ProtiGeno, a {\bf Prot}ein language model-based method for short {\bf Gen}e predicti{\bf o}n (Figure~\ref{fig:fig1}). ProtiGeno can be applied, without any further retraining, to find short genes in any prokaryotic species. We systematically compared ProtiGeno with baseline gene finders on short coding and noncoding regions from 4,288 prokaryotic genomes. ProtiGeno consistently outperforms the baseline gene finders in terms of accuracy, recall, and F1 score with a lower precision rate (Table~\ref{tab:full_result_table_1}). Our main contributions are: 

\begin{itemize}
    \item We develop a gene parsing pipeline that extracts protein sequences for short coding and noncoding regions.
    \item We collect a database of ORFs from 4,288 genomes, consisting of 145,232 short coding regions and 3,465,408 noncoding regions extracted using our pipeline.
    \item We develop ProtiGeno, a deep learning-based short gene finder using protein language models. ProtiGeno opens up new opportunities for discovering previously uncharacterized short genes and can expand our knowledge of prokaryotic organisms.
\end{itemize}

\begin{table*}
\label{tab:full_result_table_1}
\centering
\begin{tabular}[c]{lcccc}
\hline
Method & Accuracy & Precision & Recall & F1 \\
\hline
ProtiGeno & \textbf{0.933} $\pm{0.009}$ & 0.930 $\pm{0.019}$ & \textbf{0.928} $\pm{0.012}$ & \textbf{0.929} $\pm{0.009}$ \\
\hline
GeneMarkS (PGP) & 0.910 $\pm{0.006}$ & \textbf{1.0} $\pm{0.000}$ & 0.809 $\pm{0.012}$ & 0.894 $\pm{0.007}$ \\

GeneMarkS (GP) & 0.923 $\pm{0.005}$ & \textbf{1.0} $\pm{0.000}$ & 0.837 $\pm{0.011}$ & 0.911 $\pm{0.007}$  \\
\hline
Prodigal (PGP) & 0.877 $\pm{0.006}$ & \textbf{1.0} $\pm{0.000}$ & 0.741 $\pm{0.012}$ & 0.851 $\pm{0.008}$ \\

Prodigal (GP) & 0.892 $\pm{0.005}$ & \textbf{1.0} $\pm{0.000}$ & 0.772 $\pm{0.011}$ & 0.872 $\pm{0.007}$\\
\hline
FragGeneScanRS (PGP) & 0.790 $\pm{0.006}$ & \textbf{1.0} $\pm{0.000}$ & 0.555 $\pm{0.011}$ & 0.714 $\pm{0.009}$ \\

FragGeneScanRS (GP) & 0.813 $\pm{0.005}$ & \textbf{1.0} $\pm{0.000}$ & 0.604 $\pm{0.010}$ & 0.753 $\pm{0.008}$\\
\hline
Random & 0.499 $\pm{0.003}$ & 0.472 $\pm{0.002}$ & 0.501 $\pm{0.004}$ & 0.486 $\pm{0.002}$ \\
\end{tabular}
\caption{Comparison of ProtiGeno with baseline gene finders. We randomly split the genomes into ten folds. We evaluate the baselines based on two measures. PGP stands for Precise Gene Prediction and counts true positives when the 3' and 5' ends match. GP stands for Gene Prediction and counts true positives when the 3' end matches with a possible misplacement of the 5' end. We report the mean and standard error of the mean (SEM) of the test performance across ten folds. Due to high computational cost, we did not include Balrog results as each genome takes approximately 15 minutes to process, which amounts to 45 days on all 4,288 genomes}
\end{table*}%

\section{Data collection and model architecture}

{\bf Short prokaryotic gene data.} We collected 4,288 prokaryotic genomes from the NCBI GenBank database and extracted all short ORFs of size between 87 and 177 nucleotides (nts) annotated from protein homology or RefSeq ~\cite{pruitt2007ncbi}. This data comprised 145,232 short protein-coding regions and 3,465,408 short noncoding regions (see Appendix I for more details regarding data parsing and filtering). The histogram of the number of ORFs across the genomes (Figure \ref{fig:coding vs. noncoding counts}) shows that, on average, each prokaryotic genome has 34 coding and 808 noncoding regions, ranging from genomes having as low as two coding and no noncoding regions to genomes having as high as 436 coding and 6,227 noncoding regions. The data is highly skewed towards noncoding regions (1:24 ratio). In order to train a balanced-class classifier in ProtiGeno, we sampled the same number of noncoding regions as the number of coding regions. Also, to control for the ORF length as a potential spurious predictive feature, we sampled the noncoding regions from the same length distribution as the coding regions per genome (see Appendix \ref{alg:sampling_algorithm}). We also removed all duplicate coding and noncoding regions to avoid bias and overfitting, resulting in 128,578 unique short coding and 143,325 noncoding regions.

{\bf Protein language model and neural network classifer.} We utilize ESM-1b~\cite{rives2021biological}, a masked protein language model, to embed the translated open reading reading frames. ESM-1b is a 33-layer self-supervised transformer model with 650M parameters trained on the UniRef50 dataset ~\cite{suzek2015uniref}. We used ESM-1b to embed short genes into embedding vectors of size 1280. We then trained a $7$-layer fully connected deep neural network to capture the non-linearities in the protein embeddings (see Appendix~\ref{tab:ablation_study} for ablation studies and details on the network). 

\section{Results}

{\bf Gene prediction performance across all prokaryotic genomes.} To systematically test the performance of ProtiGeno on all the 4,288 genomes, we do a 10-fold cross-validation, where we use 90\% of the genomes for training and the other 10\% for evaluation. This way of splitting the data avoids data leakage across the genomes. In Table \ref{tab:full_result_table_1}, we report the mean and standard error of the mean (SEM) of test accuracy, precision, recall, and F1 score across folds for ProtiGeno, GeneMarkS ~\cite{besemer2001genemarks}, Prodigal ~\cite{hyatt2010prodigal} and FragGeneScanRS ~\cite{van2022fraggenescanrs} (see Appendix \ref{tab:full_10_fold_validation_results} for detailed results). All baseline methods are run with default settings and no hyperparameter tuning is involved. The performance metrics were averaged over all the ORFs in the genomes of the test fold. For baselines, we report results using two measures. PGP stands for Precise Gene Prediction and counts true positives when the 3' and 5' ends match. GP stands for Gene Prediction and counts true positives when the 3' end matches with a possible misplacement of the 5' end. Table \ref{tab:full_result_table_1} shows that ProtiGeno performs consistently better than the baselines in terms of accuracy, recall, and the F1 score. The higher recall rate comes at the cost of slightly lower precision. In addition to the 10-fold cross validation, we additionally report the per-genome evaluation in appendix \ref{tab:genome_wise_comparison}, where we report the percentage of genomes on which ProtiGeno outperforms the baseline methods. ProtiGeno outperforms all baseline methods on the majority of the genomes at accuracy, recall, and F-1 score, showing a similar trend as table \ref{tab:full_result_table_1}, where ProtiGeno achieves a higher recall rate at the cost of a lower precision. We analyze the false positives of ProtiGeno in more detail in the next subsections.

{\bf Gene prediction performance on representative genomes.} We tested ProtiGeno on a real-world gene prediction task. We selected five organisms identified as model organisms (MOs) in ~\citet{dimonaco2022no} due to their diversity in genome size, GC content, and their quality and comprehensiveness of genome assembly and annotation. These organisms include (1) \textit{Bacillus subtilis subsp. subtilis str. 168} (2) \textit{Caulobacter vibrioides NA1000} (3) \textit{Escherichia coli str. K-12 substr. MG1655} (4) \textit{Mycoplasmoides genitalium G37}, and (5) \textit{Staphylococcus aureus subsp. aureus NCTC 8325}. We report the precision and recall for ProtiGeno against four baselines, including GeneMarkS ~\cite{besemer2001genemarks}, Prodigal ~\cite{hyatt2010prodigal}, FragGeneScanRS ~\cite{van2022fraggenescanrs}, and Balrog ~\cite{sommer2021balrog} in Table \ref{tab:Table2-gene-prediction}. ProtiGeno achieves a higher recall rate than the baselines with a significant margin at a cost of a drop in precision (higher overal F1 score).

\begin{table*}
    \centering
    \begin{tabular}{lccccccccccc}
        \toprule
        Genomes & \multicolumn{2}{c}{\centering ProtiGeno} & \multicolumn{2}{c}{\centering GeneMarkS} & \multicolumn{2}{c}{\centering Prodigal} & \multicolumn{2}{c}{\centering FGS} & \multicolumn{2}{c}{\centering Balrog} \\
        \cmidrule(l){2-3}\cmidrule(l){4-5}\cmidrule(l){6-7}\cmidrule(l){8-9}\cmidrule(l){10-11}\cmidrule(l){12-12}
        & Prec. & Recall & Prec. & Recall & Prec. & Recall & Prec. & Recall & Prec. & Recall \\
        \cmidrule(l){1-12}
        \textit{B. Subtilis} & 0.841 & \textbf{0.736} & \textbf{1.0} & 0.601 & \textbf{1.0} & 0.615 & \textbf{1.0} & 0.490 & \textbf{1.0} & 0.543\\
        \textit{C. Vibrioides} & 0.866 & \textbf{0.587} & \textbf{1.0} & 0.430 & \textbf{1.0} & 0.446 & \textbf{1.0} & 0.273 & \textbf{1.0} & 0.149\\
        \textit{E. Coli} & 0.772 & \textbf{0.661} & \textbf{1.0} & 0.426 & \textbf{1.0} & 0.566 & \textbf{1.0} & 0.295 & \textbf{1.0} & 0.598\\
        \textit{M. Genitalium} & 0.889 & \textbf{1.0} & \textbf{1.0} & 0.375 & \textbf{1.0} & 0.625 & \textbf{1.0} & 0.500 & \textbf{1.0} & 0.625\\
        \textit{S. Aureus} & 0.909 & \textbf{0.310} & \textbf{1.0} & 0.243 & \textbf{1.0} & 0.243 & \textbf{1.0} & 0.235 & \textbf{1.0} & 0.119\\
        \bottomrule
    \end{tabular}
    \caption{Evaluation results of 5 representative organisms on ProtiGeno against four baseline gene annotators using Precise Gene Prediction (PGP): ProtiGeno consistently achieves better recall but lower precision across all five genomes, illustrating a tradeoff between ProtiGeno and the baseline gene annotators: optimizing for higher recall comes at the cost of a lower precision. ProtiGeno outputs more positive predictions, which reduces the possibility of missing a short coding region but also results in a higher false positive rate, thereby lowering precision. However, this could also potentially help us identify uncharacterized new genes.}
    \label{tab:Table2-gene-prediction}
\end{table*}

\begin{figure}[!t]
    \centering
    \includegraphics[width=0.5\textwidth]{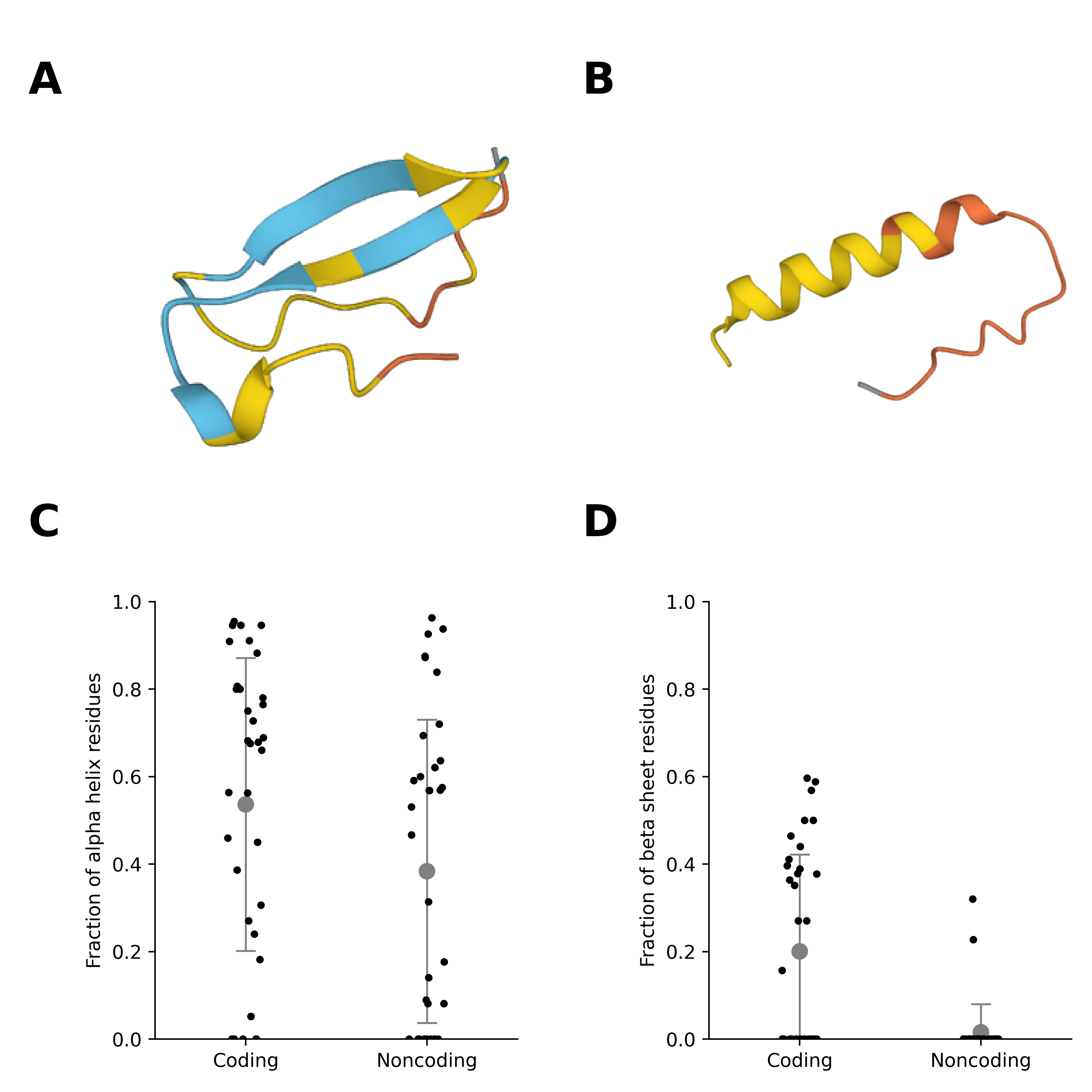}
    \caption{Structural analysis of 70 frames predicted by ProtiGeno using ESMFold. A) Plot A shows the 3D structure of a short protein coding region from sample organism \textit{F. frigiditurris}, predicted by ESMFold. B) Plot B shows the 3D stucture of a short protein noncoding region from sample organism \textit{F. frigiditurris} predicted by ESMFold. C) We compare the fraction of residues predicted as alpha helices for coding and noncoding regions across four organisms, two performing well in ProtiGeno (i.e., \textit{B. choerinum, and M. bovirhinis}) and two performing poorly in ProtiGeno (i.e., \textit{F. frigiditurris, and T. geofontis OPF15}). Two-sided t-test shows no significant difference between the means (p $>$ 0.06). D) We compare the fraction of residues predicted as beta sheets. Two-sided t-test shows significant differences between the means (p $<$ $10^{-5}$). Our analysis demonstrate that fraction of residues in beta sheets is one of the explainable predictive feature in ProtiGeno. }
    \label{fig:error_analysis_figure}
\end{figure}

{\bf Explaining ProtiGeno}. To obtain further details into ProtiGeno predictions, we performed a structural analysis using ESMFold (see Figure~\ref{fig:error_analysis_figure}). We compared the distribution of alpha helices and beta sheets between coding and noncoding regions, hypothesizing that the distribution of one is proportional to the probability of being a gene. We sampled four organisms, two performing well in ProtiGeno (i.e., \textit{B. choerinum, and M. bovirhinis}) and two performing poorly in ProtiGeno (i.e., \textit{F. frigiditurris, and T. geofontis OPF15}). We analyzed 70 coding and noncoding regions. Each data point in Figure 2C represents the fraction of alpha helix residue and Figure 2D represents the fraction of beta sheet residue. Figures show the distribution of beta sheets is significantly different between the coding and noncoding regions (p$<10^{-5}$). However, the difference is not significant for alpha helices (p$>0.06$). The analysis demonstrates that the fraction of residues in beta sheets is an explainable predictive feature in ProtiGeno. Simultaneously, alpha helix can be a potential explainable cause of the false positives.

\section{Discussion}

Tables \ref{tab:full_result_table_1} and \ref{tab:Table2-gene-prediction} show that ProtiGeno has a lower precision rate compared to the baseline methods, indicating higher false positives. We identify two factors that could potentially cause this phenomenon: (1) alpha helices confusing coding and noncoding regions and (2) the discrepancies between AlphaFold2 and ESMFold secondary structure predictions. Figure \ref{fig:error_analysis_figure} demonstrates the possibility of ProtiGeno using secondary structures as a distinguishing feature. Although the fraction of residues predicted as beta sheets by ESMFold are significantly different between coding regions and noncoding regions and can serve as a distinguishing feature, the same claim does not hold true for residues predicted as alpha helices by ESMFold (see Figure \ref{fig:error_analysis_figure}). Therefore, alpha helices can potentially cause false positives in ProtiGeno, resulting in a lower precision rate. More analysis is needed to fully understand the structural features in ProtiGeno.

In addition, our comparisons of the predicted secondary structures of 84 false positive samples from 11 genomes using AlphaFold2 and ESMFold demonstrate that the fraction of predicted secondary structures is correlated with R$^2=0.74$ (Appendix \ref{fig:esmfold-vs-AlphaFold2}). However, we still observe discrepancies between the predicted 3D structures from AlphaFold2 and ESMFold, which could potentially lead to ProtiGeno misclassifying noncoding regions as coding regions. Nonetheless, while AlphaFold2 is generally the more accurate protein structure predictor ~\cite{bertoline2023before}, the family of ESMFold methods (e.g., ESM-1b)  might be preferred for large-scale gene finding tasks due to its massive computational advantages (60 times faster), offering opportunities to identify uncharacterized short genes and novel functions in a more efficient manner.

\bibliography{tmlr}
\bibliographystyle{tmlr}

\newpage
\appendix
\onecolumn

\section{Coding region vs. all noncoding region lengths distribution before sampling \& filtering}

\begin{figure}[H]
    \centering
	\includegraphics[width=1.0\textwidth]{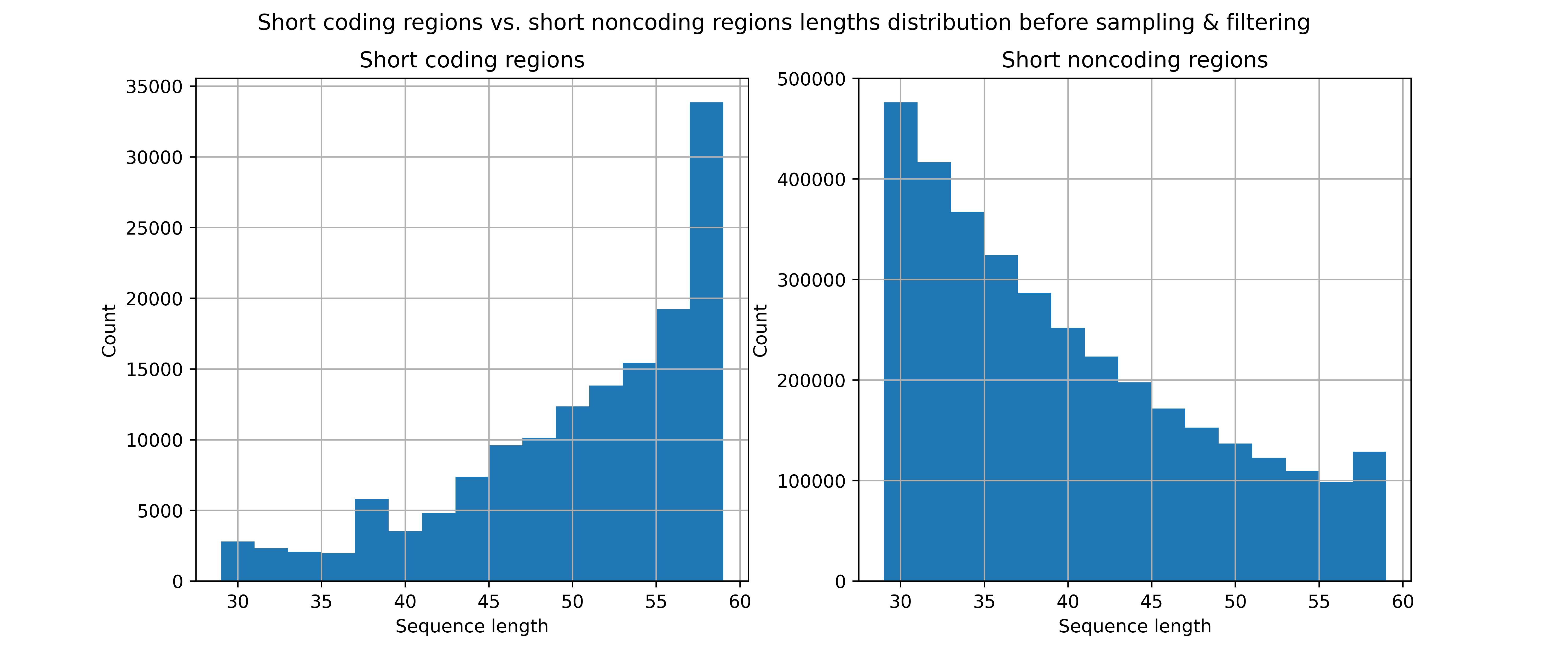}
 \caption[length distribution]{We plot the length distribution for the coding regions (in number of amino acids) as well as all noncoding regions before sampling. All short coding regions and noncoding regions are within 87 and 177 nts. The length distribution for short coding regions are skewed to the left and the length distribution for noncoding regions are skewed to the right.}
	\label{fig:coding vs. noncoding length distribution before}
\end{figure}

We present the length distribution of short coding regions and short noncoding regions before sampling and filtering. As we can see in Figure \ref{fig:coding vs. noncoding length distribution before}, the length distribution for short coding regions are skewed to the left, most of which are concentrated at 177 nts, while the length distribution for short noncoding regions are skewed to the right, most of which are concentrated at 77 nts. 

\section{Noncoding region sampling algorithm}

To avoid bias against length and overfitting in the gene classifier, we designed a sampling algorithm which samples noncoding regions from the same length distribution as the coding regions per genome.

\begin{algorithm}[H]

\label{alg:sampling_algorithm}
\begin{algorithmic}[1]
\Procedure{SampleNoncodingSeqs}{genome, CodingSeqs, NoncodingSeqs}

\State SampledNoncodingSeqs $\gets$ []
\State CodingSeqFrequencyDict $\gets$ Create frequency Dict for coding region lengths

\State NoncodingSeqFrequencyDict $\gets$ Create frequency Dict for noncoding regions lengths

\State carry $\gets$ 0

\For {key in CodingSeqFrequencyDict keys sorted in reverse order}
\State NumCodingSeqs $\gets$ CodingSeqFrequencyDict[key] \Comment{Coding Frequency Dict value for this key}

\State NumNoncodingSeqs $\gets$ NoncodingSeqFrequencyDict[key] \Comment{Noncoding Frequency Dict value for this key}

\If{NumCodingSeqs + carry $\le$ NumNoncodingSeqs}
\State Noncoding samples $\gets$ Sample NumCodingSeqs + carry samples from NoncodingSeqs
\State carry $\gets$ 0
 \Comment {If we have more noncoding samples available, we randomly sample NumCodingSeqs + carry many samples from NoncodingSeqs with length equal to key and assign carry to 0}
 \EndIf

\If{NumCodingSeqs + carry $ > $ NumNoncodingSeqs}
    \State Noncoding samples $\gets$ Sample all NumNoncodingSeqs number of noncoding regions
    \State carry $\gets$ NumCodingSeqs + carry - NumNoncodingSeqs \Comment {If NumCodingSeqs + carry is greater than the number of noncoding regions available, we simply take all NumNoncodingSeqs number of NoncodingSeqs with length equal to key and assign carry to the difference between the two numbers}
\EndIf
\EndFor

\State SampledNoncodingSeqs $\gets$ SampledNoncodingSeqs + Noncoding samples
\State \textbf{return} CodingSeqs, SampledNoncodingSeqs 
\EndProcedure
\caption{Noncoding Region Sampling Algorithm}

\end{algorithmic}
\end{algorithm}

\section{Coding region vs. noncoding region lengths distribution after sampling \& filtering}

After applying the sampling algorithm from Appendix \ref{alg:sampling_algorithm}, we also filtered out duplicate sequences to avoid bias in the dataset. As a result, we obtained a total of 128,578 unique coding regions and 143,325 unique noncoding regions, which we use for training with approximately the same length distribution, which we use to train the gene classifier. 

\begin{figure}[h]
    \centering
	\includegraphics[width=1.0\textwidth]{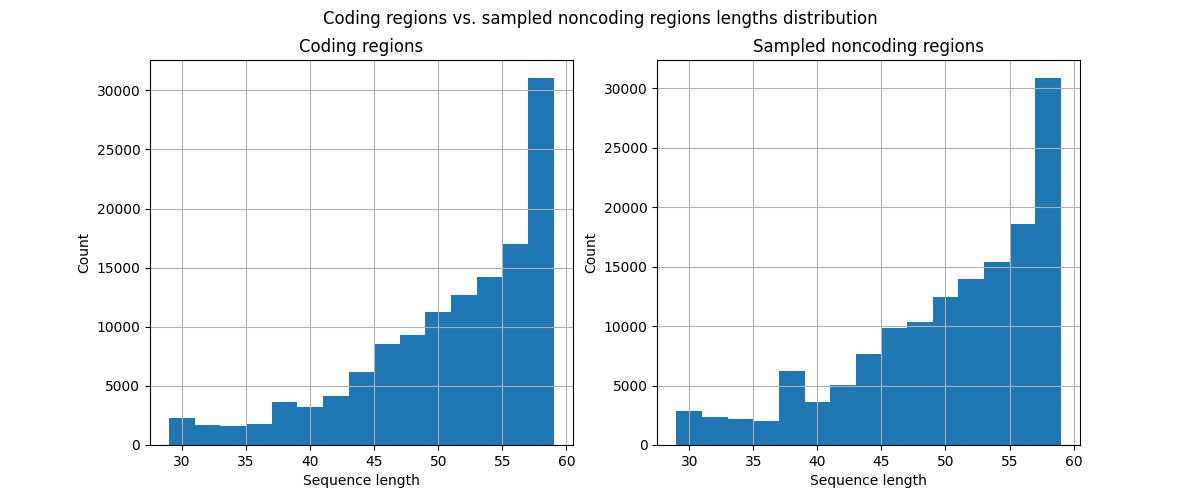}
 \caption[length distribution]{We plot the length distribution for the coding regions (in number of amino acids) as well as the sampled noncoding regions. Our sampling algorithm samples approximately the same number of noncoding regions as there are coding regions while ensuring similar length distribution to avoid bias against length in model training. }
	\label{fig:coding vs. noncoding length distribution after}
\end{figure}

\section{ProtiGeno 10-fold cross validation full results (aggregated)}
\label{tab:full_10_fold_validation_results}
As illustrated in Table \ref{tab:full_result_table_1}, we conducted a 10-fold cross validation where we train on 90\% of the genomes and evaluate on the rest of the 10\% of the genomes. Here, we present the full 10-fold cross validation results. The (aggregated) results are calculated by concatenating all predictions and groundtruths in one fold. The mean value accuracy, precision, recall and F1 score take the average across all 10 folds. 

\begin{table*}[h]
\centering
\begin{tabular}[t]{lcccccc}
\hline
\textbf{Fold} & \textbf{Train seqs} & \textbf{Test seqs} & \textbf{Accuracy} & \textbf{Precision} & \textbf{Recall} & \textbf{F1}\\
\hline
1 & 245544 & 26359 & 0.936 & 0.950 & 0.912 & 0.931\\
\hline
2 & 243767 & 28136 & 0.929 & 0.931 & 0.917 & 0.924\\
\hline
3 & 241748 & 30155 & 0.928 & 0.927 & 0.922 & 0.924\\
\hline
4 & 244110 & 27793 & 0.939 & 0.949 & 0.921 & 0.935\\
\hline
5 & 245386 & 26517 & 0.934 & 0.932 & 0.930 & 0.931\\
\hline
6 & 244359 & 27544 & 0.940 & 0.921 & 0.954 & 0.937\\
\hline
7 & 246605 & 25298 & 0.935 & 0.920 & 0.944 & 0.932\\
\hline
8 & 244532 & 27371 & 0.935 & 0.942 & 0.919 & 0.931\\
\hline
9 & 245299 & 26604 & 0.943 & 0.946 & 0.933 & 0.940\\
\hline
10 & 245777 & 26126 & 0.909 & 0.882 & 0.931 & 0.906\\
\hline
Mean (Aggregate) & - & - & 0.933 $\pm{0.009}$ & 0.930 $\pm{0.019}$ & 0.928 $\pm{0.012}$ & 0.929 $\pm{0.009}$ \\
\hline
\end{tabular}
\caption{ProtiGeno 10-fold cross validation full result (aggregated). We report the aggregated mean and standard error of mean (SEM) for accuracy, precision, recall and F1 score across 10-folds. Results are aggregated, where accuracy, precision, recall and F1 scores are calculated by concatenating predictions and groundtruths across all genomes in a fold.}
\end{table*}%

\section{Gene classifier model architecture ablation study}

We conducted an ablation study comparing the effect of model architecture on the performance of gene classifier. We trained gene classifiers with 3 model architectures: (1) 4-layer fully connected net, (2) 7-layer fully-connected net, and (3) 9-layer fully-connected layer. As shown in the table below, the 7-layer fully-connected net achieves the highest accuracy, recall, and F1 score. The 4-layer FC-Net has dimensions [1280, 32, 64, 1]. The 7-layer FC-Net has dimensions [1280, 32, 64, 128, 64, 32, 1]. The 9-layer FC-Net has dimensions [1280, 128, 256, 512, 256, 128, 64, 32, 1]. All models use xavier weight initialization and ReLU activation and are trained for 300 epochs using a learning rate of 0.01 and binary cross entropy loss. 

\begin{table*}[h]
\centering
\label{tab:ablation_study}
\begin{tabular}[t]{lcccc}
\hline
\textbf{Network Architecture} & \textbf{Accuracy} & \textbf{Precision} & \textbf{Recall} & \textbf{F1}\\
\hline
4-layer fully-connected net & 0.934 & \textbf{0.952} & 0.904 & 0.927\\
\hline
7-layer fully-connected net & \textbf{0.936} & 0.950 & \textbf{0.912} & \textbf{0.931}\\
\hline
9-layer fully-connected net & 0.925 & 0.929 & 0.911 & 0.920\\
\hline
\end{tabular}
\caption{ProtiGeno 10-fold cross validation result (aggregated): We conduct an ablation study to compare performance across 3 different neural network architectures. We evaluate these 4 models on the first fold and report the accuracy, precision, recall and F1 score.}
\end{table*}%

\vspace*{2cm}
\section{Recall vs. Coding region sequence length on GeneMarkS}

We plot the GeneMarkS PGP recall rates on coding regions of different lengths. The recall rate drops as the coding region length increases. These recall rates are different from those presented in Table \ref{tab:full_result_table_1}, where we filtered out the incomplete genes, pseudogenes and duplicate genes. 

\begin{figure}[h]
    \centering
	\includegraphics[width=0.90\textwidth]{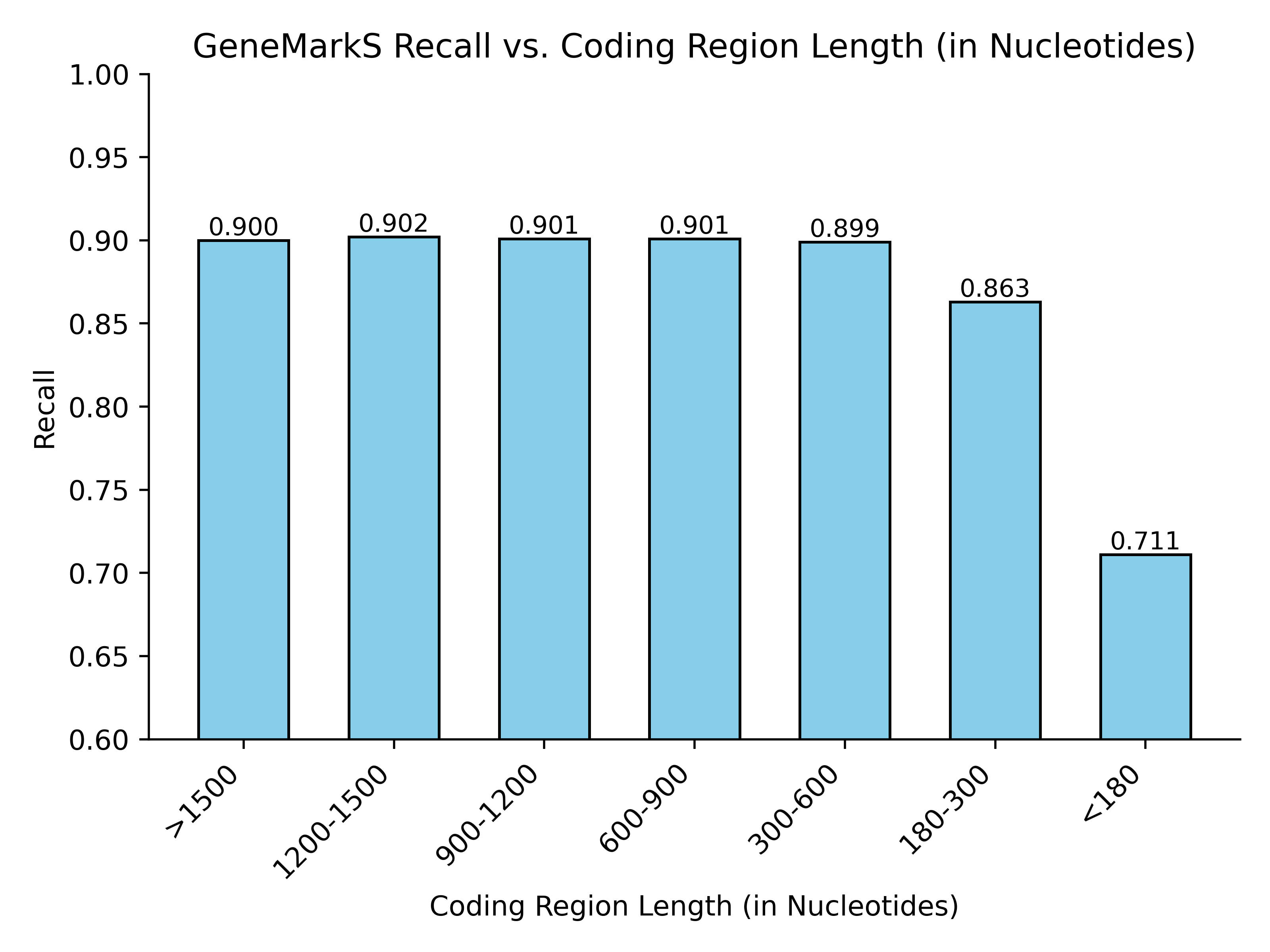}
 \caption[GMS recall vs. seq length]{GeneMarkS PGP recall rates on coding regions of different lengths. There is a downward trend in recall as the coding region length decreases, demonstrating the need to improve gene prediction performance on shorter genes. These recall values are different from those in Table \ref{tab:full_result_table_1}, where we filtered out incomplete genes, duplicate genes, and duplicate genes for training ProtiGeno. }
	\label{fig:gms_recall_vs_length}
\end{figure}

\clearpage
\section{Number of sequences vs. genome before sampling \& filtering}

We plot the number of coding regions and noncoding regions (sorted in descending order) per genome. As shown in the figure, the number of short noncoding regions exceeds the number of short coding regions by a large margin. On average, each prokaryotic genome has 34 coding and 808 noncoding regions, ranging from genomes having as low as 2 coding and 0 noncoding regions to genomes having as high as 436 coding and 6,227 noncoding regions. The data is highly skewed towards noncoding regions (1:24 ratio). To avoid class imbalance and thereby creating bias in our gene classifier, we apply a sampling algorithm described in Appendix \ref{alg:sampling_algorithm} to downsample noncoding regions. 

\begin{figure}[h]
    \centering
	\includegraphics[width=1.0\textwidth]{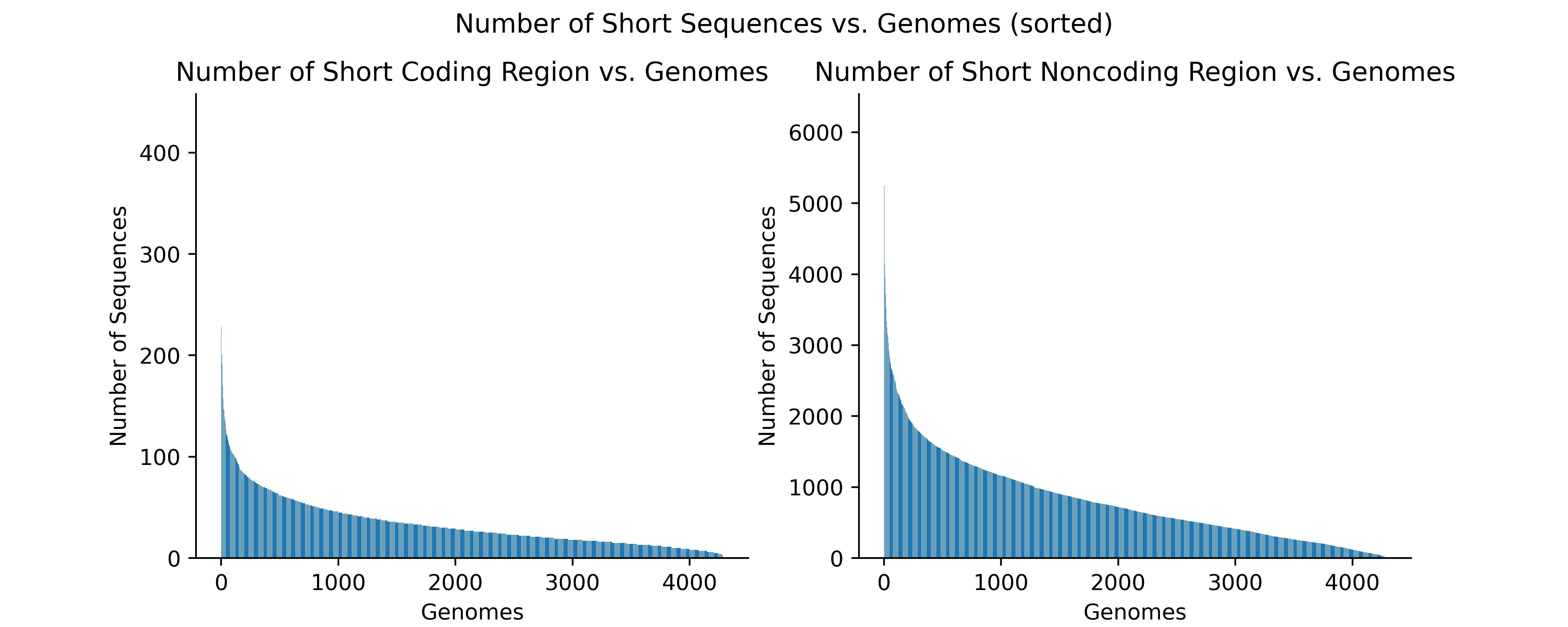}
 \caption[genome sequence counts]{Number of short coding and noncoding regions vs. genome (sorted in descending order). We plot the Number of short (between 87 and 177 NT) coding and noncoding regions for every genome, sorted in descending order. As we see that the number of short noncoding regions exceeds the number of short coding regions by a large margin, marking a significant class imbalance in our dataset. }
	\label{fig:coding vs. noncoding counts}
\end{figure}

\clearpage
\section{Fraction of secondary structure in ESMFold vs AlphaFold2. }

We plot the fraction of secondary structure in ESMFold vs AlphaFold2 .This plot is a correlation of false positives for a sample size of 11 genomes which are \textit{F. frigiditurris, T. geofontis OPF15, S. roseirectus, N. salina, M. sediminilitoris,B. wiedmannii, C. freundii, A. isosaccharinicus, N. moolapensis, A. marina, P. aryabhattai  } containing 84 sequences in total. The scatter plot represents the fraction of secondary structures predicted by the models ESMFold and AlphaFold2. The regression line on the plot shows the goodness of fit test for the data. The data points are scattered among all regions of the plot and the R$^2=0.74$ . The Welch's t-test performed for the ESMFold and AlphaFold2 groups (p$>$0.6) indicates that the null hypothesis fails to be rejected. Therefore no significant difference is observed between the predictions from ESMFold and AlphaFold2 for the false positive data. 

\begin{figure}[h]
    \centering
	\includegraphics[width=1.0\textwidth]{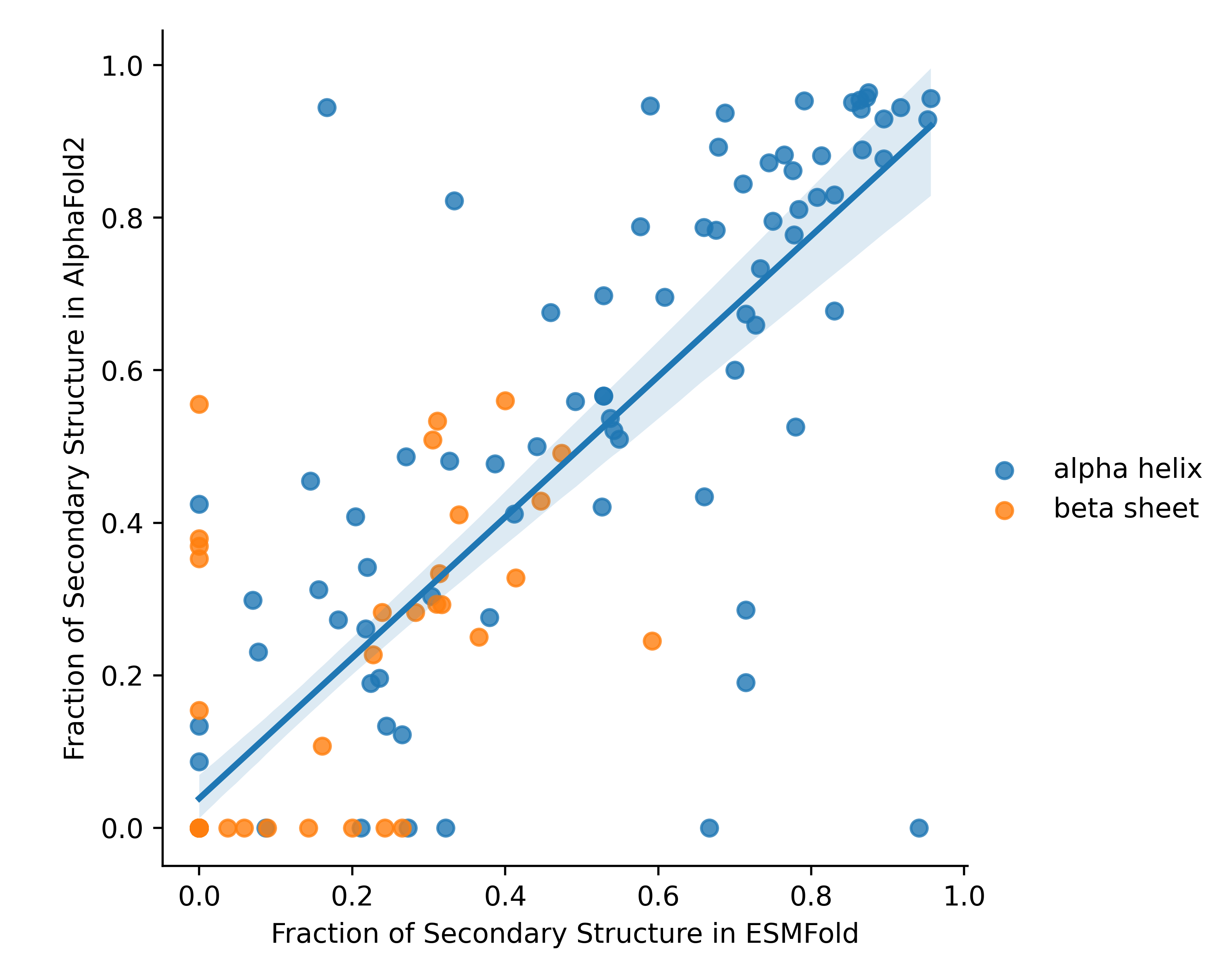}
 \caption[ESMFold vs AlphaFold2]{Correlation between secondary structure predictions by ESMFold and Aplhafold2 for 84 false positive samples from 11 genomes. The scatter plot shows the distribution of secondary structures generated by both methods. A Welch's t-test was performed to find the similarity between the predictions from the two models (p$>$0.6). The analysis fails to reject the null hypothesis, showing no significant difference between the fractions of alpha helices and beta sheets generated by ESMFold and AlphaFold2 for the false positive samples. }
	\label{fig:esmfold-vs-AlphaFold2}
\end{figure}

\section{Experimental methods}

{\bf Data collection and curation.} We collected 4,288 prokaryotic genomes from the National Center for Biotechnology Information (NCBI) GenBank database by setting the filters to (1) Prokaryotes, (2) Bacteria, (3) Archaea, (4) RefSeq, (5) Representative, (6) Reference and (7) Complete. After obtaining the genomes and annotations, we selected short coding regions with length between 87 nts and 177 nts annotated from protein homology or RefSeq ~\cite{pruitt2007ncbi}, and excluded incomplete genes and pseudogenes. We also parse noncoding short ORFs in the intergenic regions from the same genomes, select noncoding regions between 87 nts and 177 nts, and create hypothetical translations. Due to the relative abundance of intergenic regions and noncoding regions compared to the number of short genes in genomes, we sample the same number of noncoding regions as there are short genes for each genome to avoid class imbalance, which could potentially cause our classifier to over-classify the majority group (noncoding regions) due to its increased prior probability. Furthermore, to prevent our classifier model from learning input sequence length as a feature, for each genome, we applied a sampling algorithm (see Appendix \ref{alg:sampling_algorithm}) to sample the same number of noncoding regions as there are coding regions while ensuring approximately the same length distribution. In case of annotated gene duplication (two genes with the same sequence), only one protein is selected to avoid bias in the dataset. After applying our sampling algorithm and filtering out duplicates genes, we obtained 128,578 unique coding regions and 143,325 unique noncoding regions for training.

\section{Protigeno per-genome evaluation and comparison to baseline methods}
\label{sec:per-genome_evaluation}

To gain more insights on how Protigeno performs on individual genomes, we provide a comprehensive table where each genome's performance was evaluated using the model trained on the remaining 90 percent of the genomes. This table collectively summarized the model performance for all genomes and is provided \href{https://github.com/tonytu16/protigeno/tree/main/Protigeno_evaluation_results}{here} in the github repository. Note that this table consists of 4,280 genomes instead of 4,288 because after filtering out duplicate sequences, 8 genomes do not have either coding regions or non-coding regions, which we have excluded from the evaluation. In table \ref{tab:genome_wise_comparison}, we report the percentage of genomes ProtiGeno outperforms against each baseline method in terms of accuracy, precision, recall and F-1 score. Similar to table \ref{tab:full_result_table_1}, ProtiGeno outperforms the baseline methods in majority of genomes in terms of accuracy, recall and F-1 score. 

\begin{table*}[h]
\label{tab:genome_wise_comparison}
\centering
\begin{tabular}[c]{lcccc}
\hline
Baseline Method & \% of outperformed genomes (Accuracy) & \% (Precision) & \% (Recall) & \% (F1) \\
\hline
GeneMarkS (PGP) & 80.2 & 37.7 & 91.4 & 80.5 \\

GeneMarkS (GP) & 73.9 &	37.7 & 88.5 & 74.1 \\
\hline
Prodigal (PGP) & 91.6 &	37.9 & 97.4 & 92.0\\

Prodigal (GP) & 87.9 &	37.8& 96.1 & 88.2\\
\hline
FragGeneScanRS (PGP) & 93.5 & 37.9	& 99.5 & 98.9\\

FragGeneScanRS (GP) & 91.5& 37.9 & 99.3 & 98.2\\
\hline
\end{tabular}
\caption{Genome-wise comparison of ProtiGeno with baseline gene finders. We randomly split the genomes into ten folds to avoid data leakage across genes in the same genome. We report the percentage of genomes which ProtiGeno outperforms against each baseline method in terms of accuracy, precision, recall and F-1 score. Similar to the aggregated results reported in Table \ref{tab:full_result_table_1}, ProtiGeno outperforms the baseline methods in majority of genomes at accuracy, recall and F-1 score. }
\end{table*}%

\section{Computational Resources and Runtime}

Our pipeline uses sequences embeddings generated with code provided by ~\citet{dallago2021flip}. All sequences are processed using NVIDIA RTX A6000 GPU. We randomly sample 30 genomes from all 4,288 genomes we used in this paper and report their overall runtime. Detailed runtime information for the coding regions can be found \href{https://github.com/tonytu16/protigeno/blob/main/coding_regions_runtime/runtime.csv}{here}, and noncoding regions can be found \href{https://github.com/tonytu16/protigeno/tree/main/noncoding_regions_runtime/runtime.csv}{here}. Out of the 30 sampled genomes, each genome has 38 coding regions and 836 noncoding regions on average, and takes 31.4 seconds to process on average. The multi-layer perceptron models are trained on 2.3 GHz 8-Core Intel Core i9 CPU, each model used in the 10-fold cross validation in Table~\ref{tab:full_result_table_1} tables about 1 hour to train. 

\section{Baseline Implementaion and Parameters }

In this paper, we compared ProtiGeno with four baseline methods: GeneMarkS2 \cite{besemer2001genemarks}, Prodigal ~\cite{hyatt2010prodigal}, Balrog \cite{sommer2021balrog}, and FragGeneScanRS \cite{van2022fraggenescanrs}. All of these baseline methods were run on whole prokaryotic genomes. The baseline methods were all run with no specific hyperparameter tuning. All of them were ran with their default input options. For GeneMarkS2, we mentioned bacteria as the genome type, gcode was set to 11 since that is the default code for bacteria, and the output file format was specified as gff3. In Prodigal's case, the input file's protein translation was specified as prot\_trans, and the output file format was gff. Balrog was run with its default parameters, as mentioned in their colab notebook; no changes were made beyond uploading the whole genome as input. Similarly, FragGeneScanRS was run with default setting without any hyperparameter tuning. 

\section{ProtiGeno evaluation results on original dataset}

In addition to the results we obtained in Table \ref{tab:full_result_table_1}, we also trained and evaluated ProtiGeno on ALL available coding regions and non-coding regions embeddings without any sampling, considering that in real-world scenarios, noncoding ORFs are much more abundant. We split genomes into ten folds with the same randomness as Table \ref{tab:full_result_table_1} and report the mean and standard error of the mean (SEM) in Table \ref{tab:genome_wise_comparison}. As we see that, compared to Table \ref{tab:full_result_table_1}, ProtiGeno still outperforms all baseline models on recall, indicating the method is less likely to miss real coding regions compared to the existing baseline methods, although at the cost of a lower precision. Notice that the precision is much lower than those reported in Table \ref{tab:full_result_table_1} because with the majority of the training and testing data being noncoding regions, ProtiGeno creates more false positives compared to the sampled, balanced data.

\begin{table*}[h]
\label{tab:all_genome_evaluation}
\centering
\begin{tabular}[c]{lcccc}
\hline
Method & Accuracy & Precision & Recall & F1 \\
\hline
ProtiGeno & 0.935 $\pm{0.028}$ & 0.376 $\pm{0.083}$ & \textbf{0.928} $\pm{0.012}$ & 0.529 $\pm{0.090}$ \\
\hline
GeneMarkS (PGP) & 0.993 $\pm{0.001}$ & \textbf{1.0} $\pm{0.000}$ & 0.809 $\pm{0.012}$ & 0.894 $\pm{0.007}$ \\

GeneMarkS (GP) & \textbf{0.994} $\pm{0.001}$ & 0.998 $\pm{0.000}$ & 0.837 $\pm{0.011}$ & \textbf{0.910} $\pm{0.007}$  \\
\hline
Prodigal (PGP) & 0.990 $\pm{0.001}$ & \textbf{1.0} $\pm{0.000}$ & 0.741 $\pm{0.012}$ & 0.851 $\pm{0.008}$ \\

Prodigal (GP) & 0.992 $\pm{0.001}$ & 0.998 $\pm{0.000}$ & 0.772 $\pm{0.011}$ & 0.871 $\pm{0.007}$\\
\hline
FragGeneScanRS (PGP) & 0.984 $\pm{0.001}$ & \textbf{1.0} $\pm{0.000}$ & 0.556 $\pm{0.011}$ & 0.714 $\pm{0.009}$ \\

FragGeneScanRS (GP) & 0.985 $\pm{0.001}$ & 0.996 $\pm{0.001}$ & 0.604 $\pm{0.010}$ & 0.752 $\pm{0.008}$\\
\hline
Random & 0.499 $\pm{0.003}$ & 0.472 $\pm{0.002}$ & 0.501 $\pm{0.004}$ & 0.486 $\pm{0.002}$ \\
\end{tabular}
\caption{Comparison of ProtiGeno with baseline gene finders evaluated on ALL available coding regions and non-coding regions, considering that in real-world scenarios, noncoding ORFs are much more abundant. We split all genomes into ten folds in with the same random seed as table 1 to avoid data leakage across genes in the same genome. We report the mean and standard error of the mean (SEM) of the test performance across ten folds.}
\end{table*}%

\end{document}